\documentclass[apl,twocolumn,showpacs,preprintnumbers,amsmath,amssymb]{revtex4}

\usepackage{epsfig}

\begin{document}

\title{Magnetic memory and current amplification devices using moving domain walls} 

\author{S. E. Barnes}
\affiliation{Institute for Materials Research, Tohoku University, 
Sendai 980-8577, Japan, and Physics Department, University of Miami, Coral Gables, FL 33124, USA}

\author{ J. Ieda  and  S. Maekawa}
\affiliation{Institute for Materials Research, Tohoku University, 
Sendai 980-8577, and CREST, Japan Science and Technology Agency (JST), Kawaguchi 332-0012, Japan}

\date{\today} 
\begin{abstract}
{
A moving magnetic domain wall produces an electromotive force (emf). It is therefore possible to read the state of a magnetic memory device via the  emf it produces when subject to an interrogation pulse. It is also possible to amplify currents in pulse circuits, opening up the possibility of all magnetic logic circuits.
 }
\end{abstract}


\maketitle

There is considerable interest in magnetic random access memory (MRAM) both from  the technological and fundamental levels. It is now well appreciated that the state of such a device can be switched through the angular momentum transfer process, i.e., by simply passing a suitable current\cite{smbook}. Very recently the use of domain walls as the active part of such memory has attracted a lot of attention since the switching currents $j_c$ are much smaller than those relevant to pillar designs\cite{iclist}. The theory of the angular momentum process is actually easier in this case and, apart from issues of how relaxation should enter, is now fairly well understood\cite{Slon,Berger,bazaliy98,TK,sebsm05}. While the state of such a domain wall memory device might be switched by a current, it is usual  to imagine {\it reading\/} its state using magneto-resistive effects\cite{Versluijis} . Both reading and writing a  MRAM element would involve regular semiconductor technology. The purpose of this letter is to show how such a domain wall MRAM element might be {\it read\/} by making it drive a current in an external circuit, i.e., by the inverse of the writing process. Also  discussed is how currents might be amplified using other magnetic elements which function through the displacement of domain walls. In this manner it becomes feasible to use magnetic only devices in logical circuits.

\begin{figure}[b!]\centerline{\epsfig{file=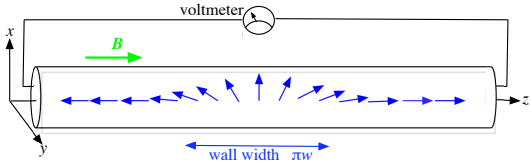,width=2.8in}  
}
\caption[toto]{
Wire geometry discussed in connection with the wall generated emf. The domain wall show at the center of the wire moves in the $z$-direction due to the pressure $P_z$ generated by the applied magnetic field $\vec B$.  The emf generated by the wall might be measured by the external voltmeter as shown.
}
\label{F0}\end{figure}

The principal  ingredient in what is discussed here is the existence of an electromotive force (emf) ${\cal E}$ generated by a moving domain wall subject to a pressure. That this exists can be shown in a number of ways. Here it will be deduced from the principles of the conservation of angular momentum and energy. Consider the wire geometry shown in Fig.~\ref{F0}. For simplicity the wire is assumed to be made of a half metal ferromagnet and to have a circular cross section. 
The anisotropy energy is such that far from the domain wall the magnetization points along the wire, here taken to be the $z$-direction. A pressure $P_z$ is exerted on the wall. Again for simplicity imagine, initially, this is due to a static  magnetic field $\vec B = B_0 \hat {\bf z}$. The pressure exerted by this field is $P_z = 2 Sg \mu_B B_0/v_c$, where $v_c$ is the unit cell volume and $S$ the magnitude of angular momentum in this volume. The basic torque transfer process\cite{Slon,Berger,bazaliy98,TK,sebsm05} follows from the conservation  of, $S_z$, the  $z$-component  of the {\it total\/} angular momentum.  An electron which has $s_z = - \hbar/2$ when it is to the left of the wall, will have $s_z = + \hbar/2$ when it is to the right. The change $\hbar$ of angular momentum is absorbed by the wall resulting its movement from $z$ to $z+\Delta z$. For the displacement  $\Delta z$ the change in angular momentum is $2S \Delta z{\cal A} \hbar / v_c$ where ${\cal A} $ is the cross-sectional area. 
Given an electrical current density $j$, using the conservation of $S_z$,  the wall velocity,
\begin{equation}
v_0 = \frac{v_c}{2eS}j \equiv C_0 j.
\label{v}
\end{equation}
This agrees with the usual result\cite{bazaliy98,TK,sebsm05} for the angular momentum (or torque) transfer mechanism.  Taking the 
lattice constant for Ni and assuming $S=1$, the constant $C_0 \approx 3 \times 10^{-11}$m$^3$/C.  Since the wall velocity $v_0$ is finite,  $(P_z {\cal A})v_0$ is the rate at which the wall is doing work. The conservation of energy dictates that this work must be given to the external circuit as a power $ {\cal E} (j {\cal A})$, i.e., the wall produces an emf,
\begin{equation}
{\cal E} = \frac{P_z v_c}{2eS},
\label{emf}
\end{equation}
again agreeing with the result obtained by other methods\cite{sebsm04}.  Accounting for the finite polarization $p$ of the conduction electrons this becomes ${\cal E} =  p (P_z v_c/2eS)$. This emf can be used to drive a current in an external circuit, in a process in which the magnetic energy stored in the wire is being converted to electrical energy.  For $p\sim 1$, ${\cal E}$ is approximately 100$\mu$V/T.

The devices described below use the emf given by Eqn.~(\ref{emf})  in an essential manner. Importantly for the present purposes, a magnetic field is not the only method by which to exert a pressure. Alternatively, use is made of the inherent energy of a domain wall. The magnetic energy per unit area of a wall $\sigma_w=  \sqrt{KA}$ where $A$ and $K$ are the exchange stiffness and magneto-crystalline anisotropy, respectively. For a wire of cross-sectional area ${\cal A}(z)$ the wall energy is $U_w = \sigma_w {\cal A}(z)$,  there is a force $ F_w = - \frac{\partial }{\partial z} U_w$ along the $z$-direction, and a pressure
\begin{equation}
P_z = - \frac{1}{{\cal A}(z)} \frac{\partial U_w}{\partial z}.
\end{equation}
A constant pressure $P_z$ and hence a constant emf $\cal E$, given by Eqn.~(\ref{emf}), corresponds to a cross-sectional area ${\cal A}(z) = {\cal A}_0 e^{-z/d}$ whence $P_z = \sigma_w /d$ and ${\cal E} =C_0 (\sigma_w/d)$.  With these expressions, the maximum values of $P_z \approx \sigma_w /w \approx  K$ and ${\cal E} \approx C_0 K$ correspond to $d \approx w$, where $w= \sqrt{A/K}$ is the wall width. For Permalloy, with $K \approx 10^5$J/m$^3$, this gives ${\cal E} \approx 3\, \mu$V. Larger values of $P_z $ and ${\cal E} $ possible when ${\cal A}(z)$ changes more rapidly, corresponding to smaller values of $d$,  require additional discussion and will not be dealt with here.

\begin{figure}[t]\centerline{\epsfig{file=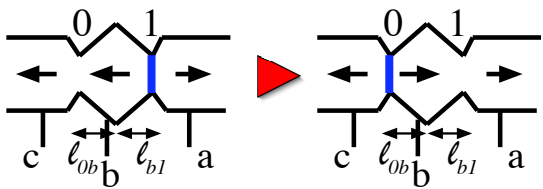,width=2.8in}  
}
\caption[toto]{A memory element which will produce a current when read. To the left, the system is in the ``1"-state. The middle contact $b$ is placed just to the left of middle wide point of the device. A current between $a$ and $b$ will carry the wall past the unstable equilibrium point. As it moves towards the ``0"-state under the influence of a $P_z$ implicit in the device shape, it will produce an output emf between $b$ and $c$. No emf occurs if the system is in the 
``0"-state. The system can be switched between ``0" and ``1" by a passing a suitably directed current between $a$ to $c$. }
\label{F1}\end{figure}

Consider, from the present perspective, the operational principle of the  basic ``peanut" memory device\cite{Versluijis} illustrated in Fig.~\ref{F1}.
The domain wall has two stable equilibrium positions $``0"$ and $``1"$ and might be switched from one to the other using an external current. Such a device can made to give a {\it current output\/} using the emf generated by the moving wall.  In such a design, the device is read by applying a current between $a$ and $b$. It is important that $b$ lies just beyond the point at which the wall energy has a maximum. If the wall lies in position $``1"$, it will be dislodged by this reading pulse, and once it passes the half-way point, will induce an output in the circuit connecting $b$ and $c$. Clearly, there will be no such current pulse if the wall is initially in position $``0"$. Here the pressure  $P_z$, switching current $j_c$ and emf reflect the shape of the bridge. 

\begin{figure}[t]\centerline{\epsfig{file=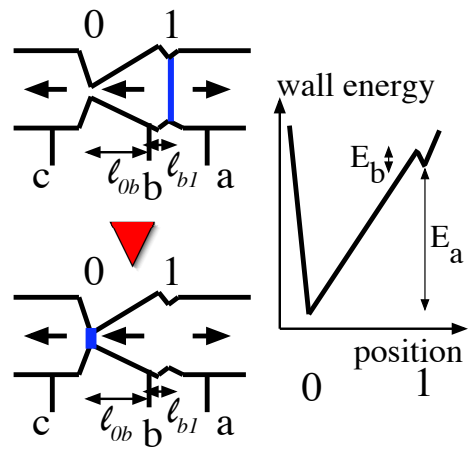,width=2.4in}  
} 
\caption[toto]{A power amplifier. The contact $b$ is again placed just to the left of a wide point. Starting with the initial state, top left, a pulse between $a$ and $b$ moves the wall from ``1" to $b$, i.e., to a point at which there is a pressure $P_z$ {\it to the left}. There is an emf between $b$ and $c$ as the wall moves between $b$ and ``0". The final state is shown at the bottom left.  To the right is shown the energy profile of the device.}
\label{F2}\end{figure}

Current and power amplification might be achieved by the device shown in Fig.~\ref{F2}. Now only the equilibrium situation $``0"$ is a truly stable. An initialization pulse or a suitably magnetic field cycle latches the system in the metastable equilibrium position $``1"$. A small short current pulse between $a$ and $b$ causes the wall to leave $``1"$ and the  pressure $P_z$ implicit in the design produces an  emf and hence a large current output in that part of the circuit which is connect to $b$ and $c$. Isolation of input and output is afforded by the small distance 1 to $b$ and again it is important that $b$ lies just beyond the point at which the wall energy has a maximum. The potential seen by the wall is also illustrated in Fig.~\ref{F2}.  A large gain implies a small barrier, of height $E_b$, at an energy $E_a$ relative to $``0"$, and which is large compared to $E_b$. In fact, the power gain $g$ is limited by, and is approximately equal to, 
\begin{equation}
g = {E_a\over E_b}.
\label{11}
\end{equation} 
The input pulse must raise the wall over the barrier, i.e.,  must give an energy $E_b$ while the maximum energy which is given to the external output circuit is evidently $E_a$. 

An advantage of the present devices is the macroscopic nature of the domain wall. This reduces enormously the effects of thermal fluctuations. With the wall in position $``1"$ of Fig.~\ref{F2} it might either tunnel or be thermally excited out of this unstable equilibrium. For the sake of illustration imagine the the effective field due to the $P_z$ is 0.1T or $\sim 0.1$K. Even if the wire only has $10^2$ atoms in both perpendicular directions there are $10^4$ spins in a cross-section and, for a wall with a length of $w \sim 10^2$ spins, it is implied that the  barrier height is $10^6 \times 0.1\sim 10^5$K which precludes the possibility of escape by both thermal excitation and tunnelling, even at room temperature. 

The present devices are inductive in nature, i.e., they store energy in magnetic form. The emf given by Eqn.~(\ref{emf}) reflects the exchange of this energy with the external world.  If the material losses were infinitesimally small, it would be possible to construct dissipation free circuits. In order to understand this, consider the memory device of Fig.~\ref{F1} with the wall at ``1".  Imagine a digital circuit  designed to operate with pulses of fixed duration $T$ and that there is an input voltage $V_{ab}$ applied across the part $ab$. The resistance between $ab$ is $R_{ab}$. The wall produces an emf ${\cal E}_i$,  Eqn.~(\ref{emf}), which by Lenz' law opposes $V_{ab}$. There is some $ab$ current $i_{ab}$ required to move the wall at the desired speed, $v_0$ and which can be deduced from Eqn.~(\ref{v}), $T$,  and the device geometry (see below).  The ratio of the dissipation ${i_{ab}}^2R_{ab}$ to $i_{ab}  {\cal E}_i$ the  power delivered to the device is ${\cal R}\equiv {i_{ab}} R_{ab}/{\cal E}_i$ and  indeed vanishes as $R_{ab}$ goes to zero and otherwise is small if ${\cal E}_i$ is large. Since $V_{ab} - {\cal E}_i = i_{ab} R_{ab}$ it follows $V_{ab} \approx {\cal E}_i$ when  ${\cal R} \ll 1$. 

Given the device dimensions, the resistivity $\rho$ and the anisotropy constant $K$, the minimal pulse length $T$, and hence the maximum operating frequency, is determined. Take the lengths $\ell_{0b} \approx \ell_{b1}$ in Fig.~\ref{F1} to be minimal, i.e., of order $w$. The voltage drop $i_{ab} R_{ab} \approx \rho j_{ab} w$ where $j_{ab}$ is the current density corresponding to $i_{ab}$.  From Eqn.~(\ref{emf}), $v_0 = C_0 j_{ab}$ while from the dimensions $v_0\approx  \ell_{b1}/T  \approx w/T$. It  follows ${\cal R} \approx \rho w^2/{C_0}^2  TK =  \rho A/{C_0}^2 TK^2$ using ${\cal E}_i \approx C_0 K$. For Permalloy with  $\pi w\sim 30$nm and $K \approx 10^5$J/m$^3$, requiring ${\cal R} = 0.2$, gives $T = 5\times 10^{-8}$s. {\it However}, ${\cal R} \propto \rho / K^{2}$ and high conductance, {\it hard}, magnetic materials are best suited to the present purpose. Permalloy is, in fact, engineered to be both extremely soft and to have a high resistance. Clearly $T$ will be many orders of magnitudes smaller for a hard magnet of low resistivity.

The output from the basic memory devices might be combined in such a fashion as to perform basic logical functions. If the voltages are maintained at a certain level ${\cal E}_i$, after a number of steps in the logic chain the current transmitted to the next stage in the cascade will be attenuated below optimal levels. At this point the amplifier of Fig.~\ref{F2} can be used to restore the current levels. Unlike usual amplifiers, the energy used in the amplification process is stored in the device by the initialization, as described above. Assume ${\cal R} \ll 1$. The operating voltage ${\cal E}_i$ is reflected in the value of $d$ associated with both parts between $1b$ and $b0$. The time $T$ and the input current $i_{ab}$ determine $\ell_{b1}$. The energy input  $\int_0^T i_{ab}\mathcal{E}_i dt \approx  E_b$ while the output current is determined by $\int_0^T i_{bc}\mathcal{E}_i dt \approx  E_a$. The emf $\mathcal{E}_i$ is designed to be time independent whence, e.g., $\int_0^T i_{ab}\mathcal{E}_i dt = \langle i_{ab} \rangle \mathcal{E}_i  T$ defines the average current $\langle i_{ab} \rangle$. The current gain is then,
\begin{equation}
\frac{\langle i_{bc} \rangle}{\langle i_{ab} \rangle} \approx \frac{E_a}{E_b} = g,
\end{equation}
i.e., corresponds to the estimate made above.
 
The perpendicular anisotropy $K_\perp$ has been assumed negligible. This is usually {\it not\/} the case for thin film wires. Square or round wires\cite{Yamamguchi06}, and other methods, must be used to reduce this anisotropy energy to a minimum.

In summary, it is possible to construct low dissipation magnetic memory and amplification devices which use in an essential way the emf generated by a moving domain wall.

This work was supported by a Grant-in-Ad for Scientific Research on
Priority Areas from the Ministry of Education, Science, Culture and
Technology of Japan, NEDO and NAREGI.



\end{document}